\documentclass[12pt]{iopart}

\usepackage{iopams}
\usepackage{graphics, graphicx}

\newcommand{\PoP}{{\it Phys. Plasmas} }

\newcommand{\PRep}{{\it Phys. Rep.} }
\begin{document}

\title[Heat conduction...]{Heat conduction in 2D strongly-coupled dusty plasmas}
\author{Lu-Jing Hou and Alexander Piel}
\address{IEAP, Christian-Albrechts Universit\"{a}t, Kiel, Germany}

\date{\today}

\begin{abstract}
We perform non-equilibrium simulations to study heat conduction in two-dimensional strongly coupled dusty plasmas. Temperature gradients are established by heating one part of the otherwise equilibrium system to a higher temperature. Heat conductivity is measured directly from the stationary temperature profile and heat flux. Particular attention is paid to the influence of damping effect on the heat conduction. It is found that the heat conductivity increases with the decrease of the damping rate, while its magnitude agrees with previous experimental measurement.
\end{abstract}

\medskip
\noindent
\textbf{Introduction:}
Recently, experiments had been carried out to study thermal conduction in two-dimensional (2D) strongly coupled dusty plasmas (SCDPs) \cite{Nunomura2005b,Nosenko2008} in both crystalline and solid/liquid mixture states, and a thermal conductivity, which is independent of temperature, was found. Although both these experiments were aimed at studying the heat conduction at an atomic (molecular) level, neither of them showed many details of microscopic processes during the heat transfer. Therefore, we conduct here non-equilibrium simulation by using Brownian Dynamics method \cite{Hou2008} to study heat transfer in 2D SCDPs in more details, serving as a supplement to real experiment.

\medskip

\noindent
\textbf{Numerical simulation:}
$N=10000$ particles are simulated in a rectangular area with periodical boundary condition in $y$ direction and confining boundary condition in $x$ direction. (More details of simulation and algorithm may be found in Ref. \cite{Hou2008}.) Particles interact with each other via pairwise Yukawa potential: $\phi(r)=(Q^2/r)\exp{(-r/\lambda_{D})}$, with $Q$, $r$ and $\lambda_{D}$ being the particle charge, interparticle-distance and screening length, respectively. The strong-coupling strength is given by $\Gamma=Q^2/(ak_{B}T)$, and the screening parameter by $\kappa=a/\lambda_{D}$, where $a=(\pi n)^{-1/2}$ is the 2D Wigner-Seitz radius with $n$ being the areal number density and $k_{B}T$ the system temperature. In addition, the damping coefficient $\gamma$ is needed to fully characterize the dynamics of the system. To simplify later discussion, we also introduce here the nominal plasma frequency $\omega_{0}=[2Q^2/(ma^3)]^{1/2}$, where $m$ is the mass of a particle. In the simulation, the screening parameter is kept constant at $\kappa=1$, as it is the most typical value found in experiment, while $\Gamma$ and $\gamma$ are varied to realize different equilibrium states and different damping rates.

Our simulation is directly mimicking recent experiments \cite{Nunomura2005b,Nosenko2008}, and is different from the usual method of non-equilibrium simulation for heat conduction \cite{Donko2004}. The system is firstly brought to an equilibrium with desired temperature ($T_{0}$) in either liquid or solid state. The melting point for $\kappa=1$ is at $\Gamma^{*}\approx 180$ \cite{Hartmann2005}, and we'll denote the corresponding temperature as $T^{*}$. Then the right half of the system ($x>0$), is heated to a higher temperature ($T_{1}$) by applying a Gaussian white noise with desired strength. The evolution of the temperature profile and also the heat flux are recorded. A steady state is approached after a substantially long period.
\begin{figure}
\centering
\includegraphics[trim=15mm 65mm 12mm 10mm,clip, width=0.8\textwidth]{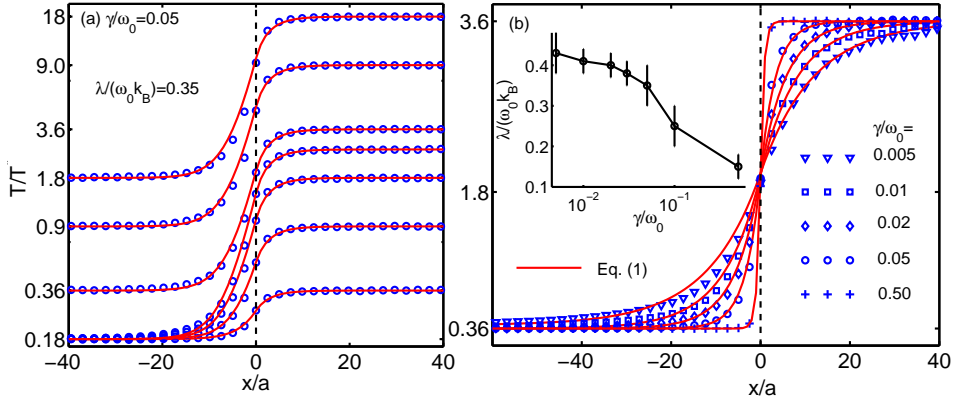}
\caption{Stationary temperature profiles for: (a) $\gamma=0.05\omega_{0}$ but different system states and temperature gradients and; (b) $T_{0}=0.36T^{*}$ and $T_{1}=3.6T^{*}$ but different damping rate. In both figures, symbols are simulation results while solid lines are fits according to Eq. (\ref{eq_Tx}). For (a), $\lambda=0.35\omega_{0}k_{B}$, and for (b) different $\lambda$ values(given in the inserted plot) are obtained.} \label{fig_T}
\end{figure}

The microscopic heat flux for $i$th particle is defined as:
$
\mathbf{J}_{i}(t)= \mathbf{v}_{i}E_{i} +\frac{1}{2}\sum^{N}_{j=1,j\neq i}\mathbf{r}_{ij}(\mathbf{F}_{ij}\cdot\mathbf{v}_{ij})- \mathbf{r}_{i}(\mathbf{F}_{ext}\cdot\mathbf{v}_{i})
$
where $E_{i}=(1/2)(mv^2_{i}+\sum^{N}_{j=1,j\neq i}\phi_{ij}) +\phi_{ext}$ is the particle energy. The total flux in a region is then a summation of the microscopic flux of all particles therein divided by its area $A$, i. e., $\mathbf{J}(t)=(1/A)\sum_{i\in A}\mathbf{J}_{i}(t)$. We are mainly interested in the $x$-component of the heat flux $J_{x}$. The three terms on the right-hand-side of above equation correspond to respectively contributions from: (1) the particle migration, which is believed to be the main mechanism of heat transport in gas and is denoted as $J_{kx}$ hereafter, (2) particle interactions, i. e., phonon scattering, which is dominant in solids and is denoted as $J_{px}$ hereafter, and (3) the external force. Since the external force acts on only a few rows of particles around the two confining boundaries, its \emph{direct} influence on the heat flux is localized. Neglecting the external contribution will bring it back to the standard one \cite{Volz2000,Lepri2003}.

\medskip

\noindent
\textbf{Analytical model:}
The heat transfer in our specific case can be described by Fourier's law: $\mathbf{J}=-\lambda\nabla{T}$ together with energy balance between heat conduction and energy dissipation due to damping: $\nabla(\lambda\nabla{T})=2\gamma n (T-T_{0})k_{B}$ \cite{Nosenko2008}, where $\lambda$ is the heat conductivity. One has, 
\begin{equation}
\raggedleft
T(x)-T_{0}=\frac{T_{1}-T_{0}}{2}e^{\sqrt{\frac{2n\gamma k_{B}}{\lambda}}x}, (x<0); \ \ \ 
T(x)-T_{1}=\frac{T_{0}-T_{1}}{2}e^{-\sqrt{\frac{2n\gamma k_{B}}{\lambda}}x}, (x>0). \label{eq_Tx}
\end{equation}
Expressions for heat flux may be obtained in a straightforward way, and we omit the results here. Since $\lambda$ is the only unknown parameter in Eq. (\ref{eq_Tx}), it may be measured by fitting the stationary temperature profile  (STP) to Eq. (\ref{eq_Tx}). It should also be mentioned that the Fourier's law could break down for low dimensional crystalline systems \cite{Lepri2003}, largely due to a slow decay of equilibrium correlations of the heat current and a divergence of the finite-size conductivity. However, both of them may be avoided in dusty plasmas because of the finite damping effect. Therefore we skip this question at this moment, while interested readers may find more discussions in \cite{Lepri2003}.

\medskip

\noindent
\textbf{Results and discussions:}
Figure \ref{fig_T} (a) shows examples of STPs for different system states and temperature gradients with $\gamma=0.05\omega_{0}$, which is close to the experimental condition of \cite{Nosenko2008}. Symbols are measurements from simulation, while solid lines are analytical fits according to Eq. (\ref{eq_Tx}). These fits give a constant heat conductivity, $\lambda=(0.35\pm 0.05)\omega_{0}k_{B}$, or in terms of thermal diffusivity $D_{T}$: $D_{T}\approx 22$mm$^2/s$ using parameters from Ref. \cite{Nosenko2008}. This value is between the experimental measurement for crystalline state ($30$mm$^2/s$) \cite{Nunomura2005b} and that for solid/liquid mixture phase ($9$mm$^2/s$) \cite{Nosenko2008}. Fits for high temperature (e. g. the two upper-most curves) suggest a slightly smaller $\lambda$. Nevertheless, the value is in the range of the error bar for the present measurement.

Figure \ref{fig_T} (b) shows STPs for different damping rate with other parameters fixed. Fits with Eq. (\ref{eq_Tx}) give a damping-dependent heat conductivity, as is shown in the inserted plot that $\lambda$ rises slightly with the decrease of $\gamma$. Note that this tendency is contradictory with that given by the analytical model in \cite{Fortov2007}, which predicts an increase of $\lambda$ with increase of damping rate and was confirmed by their experiment \cite{Fortov2007}. However, this model is based on an empirical relation between diffusion and heat conduction coefficients obtained by fitting simulation results for three-dimensional (3D) simple liquid without damping, and their experiment was also performed in a 3D dusty plasma liquid \cite{Fortov2007}. Whereas in our simulation we study 2D systems covering both liquid and solid states and/or with a liquid-solid mixture phase. So the discrepancy could have been caused by the different dimensionality and system states, as is known that transport processes depends much on these two factors. In our simulation, the damping effect is taken into account self-consistently and increase of $\lambda$ with decrease of $\gamma$ may be intuitively understood as follows. It has two effects on heat conduction: direct energy dissipation and indirect suppression of phonon propagation. The first one is only related to kinetic energy of the system, and had been explicitly taken into account in Eq. (\ref{eq_Tx}), whereas the second one affects the collective modes and is not included in Eq. (\ref{eq_Tx}). Therefore decrease of $\gamma$ means less damping of phonon propagation, more efficient heat transfer through phonon scattering and consequently a higher heat conductivity.
\begin{figure}
\centering
\includegraphics[trim=20mm 10mm 65mm 10mm,clip, width=0.75\textwidth]{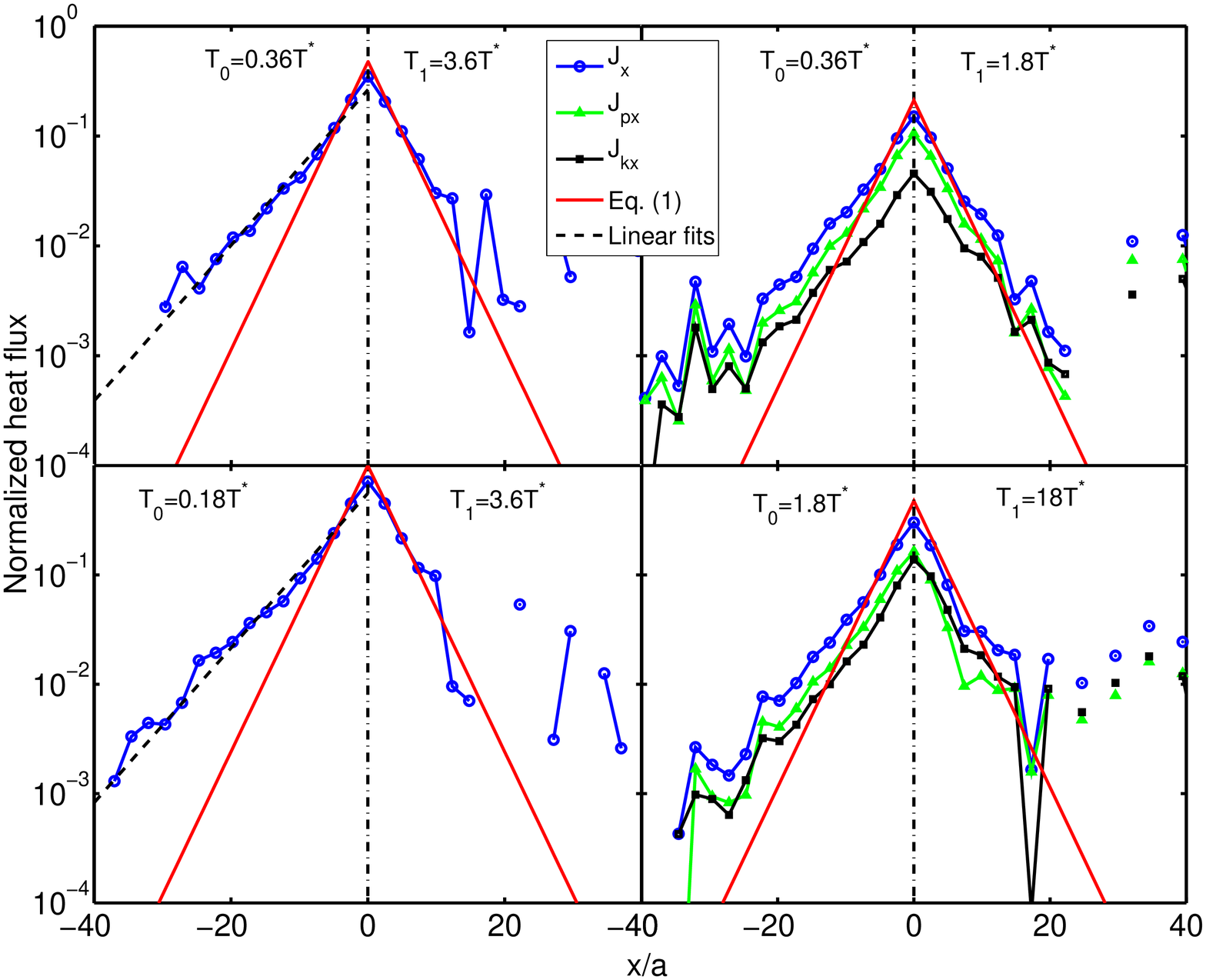}
\caption{Heat flux for different system states and different temperature gradients with $\gamma=0.05\omega_{0}$. Symbols are simulation results, solid lines are fits according to Eq. (\ref{eq_Tx}), using $\lambda=0.35\omega_{0}k_{B}$, and dash lines in the two panels on the left are direct linear fits of the heat flux with $\lambda=1.2\omega_{0}k_{B}$.} \label{fig_flux}
\end{figure}

Figure \ref{fig_flux} shows distributions of $J_{x}$, $J_{px}$ and $J_{kx}$ for different system states and temperature gradients, together with the analytical result derived from Eq. (\ref{eq_Tx}). Firstly, it may be seen that the kinetic part $J_{kx}$ and the phonon part $J_{px}$ have different weights in different system states. As expected, $J_{px}$ is clearly dominant for solid state and low temperature liquid state, while $J_{kx}$ dominates for high temperature liquid. The critical temperature where the two parts become equal is about $6T^{*}$. Secondly, the heat flux is not symmetrical about the heating interface. The decay of heat flux on the low-temperature side is slower and one needs a smaller slope, consequently a larger $\lambda$ to fit $J_{x}$ on this side, indicating a higher heat conductivity for lower temperature. Thirdly, the agreement between analytical results and simulation depends closely on system states. One generally observes a better agreement on the high temperature side and for higher temperature. These features suggest that $\lambda$ becomes temperature-dependent.

Thus, we have measured the heat conductivity $\lambda$ of 2D SCDPs by analyzing both stationary temperature profile and heat flux in non-equilibrium simulations. It is found that $\lambda$ increases with the decrease of the damping rate. In addition, our results also suggest that $\lambda$ should be temperature-dependent.

\medskip

\noindent
\textbf{Acknowledgements:}
The authors thank Prof. J. Goree and Prof. A. Melzer for valuable comments, and Dr. D. Block for valuable advices on data processing. L.J.H. gratefully acknowledges support from Alexander von Humboldt Foundation. Work at CAU is supported by DFG within SFB-TR24/A2.

\end{document}